\documentclass[12pt]{JHEP3}
\usepackage{amsmath,epsf,amssymb,latexsym,cite,graphics,bbm}

\newif\iffigs\figstrue

\DeclareFontFamily{U}{rsf}{}
\DeclareFontShape{U}{rsf}{m}{n}{
  <5> <6> rsfs5 <7> <8> <9> rsfs7 <10-> rsfs10}{}
\DeclareMathAlphabet\Scr{U}{rsf}{m}{n}
\def\O{\Scr{O}}

\def\Z{\Scr{H}}

\def\C{{\mathbb C}}

\def\P{{\mathbb P}}

\def\Z{{\mathbb Z}}

\def\Vol{\operatorname{Vol}}

\def\Gl{\operatorname{GL}}

\def\GU{\operatorname{U{}}}

\def\Hess{\operatorname{Hess}}

\def\p{\partial}

\def\cD{{\Scr D}}

\def\cL{{\Scr L}}

\def\ff#1#2{{\textstyle\frac{#1}{#2}}}

\def\zb{\bar{z}}
\def\pb{\bar{\partial}}
\def\ep{\epsilon}
\def\lambdab{\overline{\lambda}}

\def\thetab{\overline{\theta}}

\def\la{\langle}
\def\ra{\rangle}

\def\Upsilonb{\overline{\Upsilon}}

\def\psib{\overline{\psi}}

\def\phib{\overline{\phi}}
\def\gammab{\overline{\gamma}}
\def\Eb{\overline{E}}

\def\Et{{\widetilde{E}}}

\def\pit{{\widetilde{\pi}}}
\def\Phit{{\widetilde{\Phi}}}
\def\Gammat{{\widetilde{\Gamma}}}
\def\Sigmat{{\widetilde{\Sigma}}}
\def\sigmat{{\widetilde{\sigma}}}
\def\Upsilont{{\widetilde{\Upsilon}}}
\def\Vt{{\widetilde{V}}}
\def\taut{{\widetilde{\tau}}}

\def\Psib{\overline{\Psi}}
\def\Phib{\overline{\Phi}}

\def\Qb{\overline{Q}}
\def\Jb{\overline{J}}

\def\cDb{\overline{\cD}}
\def\Gammab{\overline{\Gamma}}
\def\Upsilonb{\overline{\Upsilon}}

\def\tauh{{\hat{\tau}}}
\def\Eh{{\hat{E}}}
\def\Jh{{\hat{J}}}

\def\ACO#1#2{ {\left\{ #1, #2\right\}}}

\title{Half-Twisted $(0,2)$ Landau-Ginzburg Models}
\author {Ilarion V. Melnikov and Savdeep Sethi \\
\normalsize Enrico Fermi Institute \\
\normalsize University of Chicago \\
\normalsize Chicago, IL 60637, USA\\
Email:  \email{lmel@theory.uchicago.edu,sethi@theory.uchicago.edu}
}
\abstract{We compute correlators of chiral operators in $(0,2)$ supersymmetric 
Landau-Ginzburg theories.  The class of theories and the correlators we study are relevant for extending
and testing mirror symmetry away from the $(2,2)$ locus. More generally, these methods provide $\alpha'$-exact results about certain superpotential couplings in compactifications of the heterotic string. }

\preprint{EFI-07-37}
\keywords{Superstrings and Heterotic Strings, Topological Field Theories}

\begin{document}


\section{Introduction}

The heterotic string provides a natural setting for studying  $d=4$  
Poincar\'e-invariant models of particle physics 
coupled to four-dimensional quantum gravity. A key feature of this framework is that perturbative heterotic backgrounds can be described in the RNS formalism where  the necessary and sufficient requirements to preserve  
$N=1$ space-time supersymmetry are well known: the internal world-sheet theory 
must be a unitary $c=9$ $(0,2)$ superconformal modular-invariant theory with integral
R-charges\cite{Banks:1987cy}.


Recent results \cite{Basu:2003bq,Beasley:2003fx}, building on earlier work 
\cite{Distler:1986wm, Silverstein:1995re} have shown that
contrary to initial expectations \cite{Dine:1986zy}, large classes of appropriate $(0,2)$ SCFTs may obtained as low energy
limits of $(0,2)$ supersymmetric sigma models.   In particular, $(0,2)$ deformations of many well-understood CFTs with
$(2,2)$ SUSY are unobstructed and correspond to additional massless fields in spacetime. The $(2,2)$ locus
is distinguished merely by degree of computability, in much the same way as a Gepner point is distinguished in the
moduli space of the $(2,2)$ SCFT corresponding to the quintic Calabi-Yau $3$-fold.

In light of these developments we must ask to what extent the familiar properties of $(2,2)$ theories, such as mirror symmetry, 
special geometry, and the computability of Yukawa couplings continue to hold in generic $(0,2)$ models.  This is not  idle curiosity:   
the presence of the $(0,2)$ deformations of $(2,2)$ theories indicates that even in such a familiar example as the heterotic 
string compactified on the quintic, we only understand a slice of the full moduli space!

Important progress has already been made on these issues.  Already some time ago, evidence for $(0,2)$ mirror symmetry appeared in~\cite{Blumenhagen:1996vu}. Shortly thereafter, a number of dual pairs were constructed using a $(0,2)$ analogue of the Greene-Plesser construction~\cite{Blumenhagen:1996tv, Blumenhagen:1997pp}.  The comparison of spectra in these mirror pairs utilized the techniques to compute elliptic genera of $(0,2)$ Landau-Ginzburg theories developed in~\cite{Kawai:1994qy}. More recently, mirror descriptions of $(0,2)$ gauged linear sigma models were proposed~\cite{Adams:2003zy}\ using a generalization of~\cite{Hori:2000kt}. The main object of study in these models is a truncated chiral ring that resembles the more familiar chiral rings studied in $(2,2)$ theories~\cite{Adams:2003zy, Sharpe:2005fd, Adams:2005tc}.  This ring provides a $(0,2)$ generalization of $(2,2)$ quantum cohomology~\cite{Adams:2003zy, Sharpe:2006qd}. 
In order to check these results, Katz et.~al.~\cite{Katz:2004nn, Guffin:2007mp}\ have directly computed worldsheet instanton corrections to correlators in the original $(0,2)$ linear sigma model. Significant progress has also been made in directly computing instanton generated superpotentials in geometric $(0,2)$ compactifications~\cite{Buchbinder:2002ic, Buchbinder:2002pr, Beasley:2005iu}.

There has also been substantial effort in developing a general theory of half-twisted non-linear sigma models 
\cite{Witten:2005px,Kapustin:2005pt,Tan:2006qt,Tan:2007bh}.  These theories are the natural $(0,2)$ generalizations of 
the topological sigma models familiar from the study of $(2,2)$ theories.  In worldsheet perturbation theory they 
make contact with the mathematical theory of chiral differential operators and the definition of mirror symmetry in terms of 
mirror chiral de Rham complexes.  This approach may well lead to a general framework for the study of $(0,2)$ heterotic 
compactifications and mirror symmetry off the $(2,2)$ locus.

The focus of this note is somewhat different.  We observe that in a  large class of $(2,2)$ theories, namely those
obtained as low energy limits of linear sigma models, many correlation functions are readily computable by topological
field theory techniques.  To what extent do these methods apply once the theory is deformed off the $(2,2)$ locus?
If they do extend in a straight-forward fashion, then we already have methods to compute correlators in $(0,2)$ theories.
With these computations in hand, we will be able to study the dependence of the Yukawa couplings on the bundle moduli, 
the structure of the singular locus in the CFT moduli space, the stringy resolution of the singularity, and the $(0,2)$ mirror
map.

We have studied the simplest class of such theories---$(0,2)$ deformations of gapped $(2,2)$ SUSY Landau-Ginzburg (LG) theories, 
and we have found that we can easily compute correlators of local chiral operators in the corresponding half-twisted
theory.  The correlators are given by a weighted sum over the classical supersymmetric vacua and vary smoothly across
the $(2,2)$ locus.

Given that our ultimate interest is in conformal models, it may seem strange to study gapped theories.   However, experience with 
$(2,2)$ theories suggests that the seemingly strange is natural.  For example, in Vafa's solution of topological Landau-Ginzburg 
models \cite{Vafa:1990mu}, computations are effectively done by adding a relevant deformation to the superpotential,  performing 
the necessary computations, and then taking the deformation parameters to zero.  Another relevant example is the { quantum 
restriction formula} of Morrison and Plesser \cite{Morrison:1994fr}, which relates  certain correlators in the sigma model associated 
to a Calabi-Yau hypersurface in a Fano toric variety to sums of correlators in the sigma model for  the ambient toric space.  Finally,
linear sigma model computations are often directly translatable into LG computations, either by following a dualization
procedure as in~\cite{Hori:2000kt, Adams:2003zy}, or by directly working on the Coulomb branch of the linear model \cite{Melnikov:2006kb}.

The rest of the paper is organized as follows. In section~\ref{s:defs}, we discuss $(0,2)$ deformations of $(2,2)$ linear sigma models.
This will motivate the Landau-Ginzburg theories we consider in section~\ref{s:lg}.  We compute correlators in these theories in 
section~\ref{s:semiclass} and apply the general results with a couple of illustrative examples in section~\ref{s:examples}. Section~\ref{s:conc}\ contains our conclusions.

\section{$(0,2)$ Deformations of $(2,2)$ Theories} \label{s:defs}
In this section we review the standard construction of $(0,2)$ deformations of a $(2,2)$ linear sigma model~\cite{Witten:1993yc}\ and
make some simple observations about the form of $(0,2)$ mirror symmetry.

\subsection{Basic Field Content}
The field content of the theory is easily presented in superspace.\footnote{ The reader unfamiliar with $(0,2)$ superspace will find a concise presentation
of the necessary details in \cite{Witten:1993yc}.} Parametrizing the $(0,2)$ superspace by $(x^\pm,\theta^+,\thetab^+)$, we can construct superspace 
derivatives $\cD_+,\cDb_+$ and use these to define irreducible representations of the right-moving supersymmetry.  As the linear model is
a gauge theory, it is convenient to work in Wess-Zumino gauge in presenting these multiplets. The superfields we need are as follows:
\begin{enumerate}
\item The {\em chiral}  multiplets $\Phi^i$, $i=1,\ldots,N$ are bosonic multiplets that satisfy $\cDb_+\Phi^i = 0$ and have expansion
              \[\Phi^i = \phi^i + \sqrt{2} \theta^+\psi_+^i - i \theta^+ \thetab^+ D_+ \phi^i.\] 
\item The {\em Fermi} multiplets $\Gamma^i$ are anticommuting multiplets satisfying\newline {$\cDb_+\Gamma^i~=~\sqrt{2}E^i(\Phi)$} with 
             superspace expansion
             \[\Gamma^i = \gamma_-^i - \sqrt{2} \theta^+ G^i - i \theta^+ \thetab^+D_+ \gamma_-^i - \sqrt{2} \thetab^+ E^i(\Phi).\]  
\item The {\em vector} multiplets $V_a$, $a=1,\ldots, N-d$, have the superspace expansion
             \begin{eqnarray*}
              V_{a,+}  &=&  \theta^+\thetab^+  v_+,\nonumber\\ 
              V_{a,-}   &=&   v_- - 2 i \theta^+ \lambdab_{a,-} - 2 i \thetab^+ \lambda_{a,-} + 2 \theta^+ \thetab^+ D_a.
             \end{eqnarray*}
\end{enumerate}
The ``matter'' fields $\Phi^i$ and $\Gamma^i$ are minimally coupled to the abelian vectors with charges $Q_i^a$, and the $D_+$ above
are covariant derivatives with respect to this connection.  In general, a $(0,2)$ theory need not have any natural pairing between the chiral 
and Fermi multiplets, but theories with a $(2,2)$ locus do have this pairing.  The action may then be written as
$S = S_{\text{kin}}+ S_{\text{F-I}} + S_{J}$, with
\begin{eqnarray}
\label{eq:GLSMaction}
S_{\text{kin}} & = & \int d^2 y d^2\theta \left\{ - \ff{1}{8 e_0^2} \Upsilonb_a \Upsilon_a  
- \ff{i}{2} \Phib^i (\p_-+iQ_i^aV_{a,-}) \Phi^i - \ff{1}{2} \Gammab^i \Gamma^i\right\}, \nonumber\\
S_{\text{F-I}} & = & \ff{i}{4} \int d^2y d\theta^+ (i r^a+\ff{\theta^a}{2\pi}) \Upsilon_a |_{\thetab^+ = 0} + \text{h.c.}, \nonumber\\
S_J & = & -\ff{1}{\sqrt{2}} \int d^2 y d\theta^+ \Gamma^i J_{i} (\Phi)|_{\thetab^+ =0} + \text{h.c.}.
\end{eqnarray}
Here $\Upsilon_a$ is the supersymmetric gauge-field strength, $r^a$ are Fayet-Iliopoulos parameters, $\theta^a$ are
the theta angles, and the $J_i$ are polynomials in the $\Phi^i$ with charges $-Q^i_a$.  This theory will have $(0,2)$ SUSY provided
that the $E^i$ and $J_i$ satisfy the constraint $\sum_i E^i J_i = 0$.

\subsection{Linear Model for the Quintic}
In order to construct the linear model for the quintic with $(2,2)$ SUSY, we will need six charged pairs of chiral and Fermi multiplets, with 
charges $Q_i = (-5,1,1,1,1,1)$ under a single gauge group, as well as a neutral chiral multiplet $\Sigma$.  We set $$E^i = i \sqrt{2} Q_i \Sigma \Phi^i,$$ 
and take the $J_i$ to be
\begin{equation}
J_i = \frac{\p W}{\p\Phi^i},~~~\text{where}~W = \Phi^0 P_5(\Phi^1,\ldots,\Phi^5),
\end{equation}
and $P_5$ is a generic homogeneous quintic polynomial.  The superspace presentation makes it clear that this action has $(0,2)$ 
supersymmetry.  In fact, a number of carefully made choices ensure that the action has the full $(2,2)$ supersymmetry expected 
from the heterotic compactification on the quintic with standard embedding.  

This massive theory is believed to flow to the desired superconformal theory in the IR \cite{Witten:1993yc}.  We will not repeat the many
convincing arguments for this here. For our purposes, it will be sufficient to illustrate that this might be so by considering the 
classical vacuum structure of the theory. This may be determined from the scalar potential
\begin{equation}
U = \ff{e_0^2}{2} (\sum_i Q_i |\phi^i|^2 - r )^2 + \sum_i |E^i|^2 + \sum_i |J_i|^2.
\end{equation}
The classical vacuum structure is determined by solving for $U=0$ modulo gauge equivalence. When $r$ is taken to be positive,
the first term and the gauge quotient force the $\phi^i$ to parametrize the total space of the bundle $O(-5) \to \P^4$.
The second term requires the neutral scalar $\sigma$ to be zero, while the last sum further restricts $\phi^0$ to vanish, and 
the $\phi^i$ for $i=1,\ldots, 5$ to parametrize the quintic hypersurface $P_5=0$ in $\P^4$. 

The moduli space of the $(2,2)$ conformal field theory may be paramet\-rized in terms of UV data in a simple way:  $ir+\theta/2\pi$ 
corresponds to the complexified K\"ahler class of the quintic, while the complex coefficients in $P_5$ modulo the action of $\Gl(5,\C)$ 
correspond to the $101$ complex structure moduli.\footnote{We hope the reader will forgive our usage of these geometric terms 
for the moduli of the abstract conformal field theory.  We should also point out that the linear model parametrizations of the 
K\"ahler and complex moduli are {\em not} those of special geometry.}  Of course this is not the entire story since the theory has a 
number of additional deformations that only preserve $(0,2)$ supersymmetry.

The action in eqn.~(\ref{eq:GLSMaction}) will be supersymmetric for any $E^i$ and $J_i$ that satisfy $\sum_i E^i J_i = 0$.  
When $E^i$ and $J_i$ are chosen as above, gauge invariance implies that the constraint is satisfied,  but this is not
the only way to satisfy the constraint.  In order for the conformal theory to correspond to the desired string vacuum, the deformations 
should  preserve the right-moving R-symmetry, as well as a left-moving global symmetry $\GU(1)_L$, which at the $(2,2)$ locus becomes 
the left-moving R-symmetry.\footnote{A pedagogical discussion of this point may be found in \cite{Distler:1995mi}.}   The charges of 
the fields under these symmetries are:
\begin{equation}
\begin{array}{|c|c|c|c|c|c|c|c|}\hline
\ast 		& \theta^+ & \Phi^0   & \Phi^i & \Gamma^0 & \Gamma^i & \Sigma & \Upsilon \\ \hline
\GU(1)_R  &    1         & 1 &    0     &  1      &     0               &   1       & 1 \\ \hline
\GU(1)_L 	&     0	&  1 &    0      & 0      &      -1             &   -1      & 0 \\ \hline
\end{array}
\end{equation}
It is an easy exercise to show that these symmetries fix the form of the $E^i$ and $J_0$, while for $i>0$ they constrain
$J_i = \Phi^0 G_i$, with $G_i$ arbitrary quartic polynomials in $\Phi^1,\ldots,\Phi^5$.  The $G_i$ depend on $350$ additional 
parameters, but the constraint $\sum_i E^i J_i = 0$, eliminates $126$ of these, leaving $224$ additional deformations that 
preserve $(0,2)$ supersymmetry.   These are the $224$ deformations of the tangent bundle familiar from the textbook 
discussion of the quintic \cite{GSW2}.   One lesson that is immediate from the form of the Lagrangian is that away from the 
$(2,2)$ locus all of the $J_i$ deformations, whether complex structure or bundle, ought to be treated on the same footing.

\subsection{$(0,2)$ Mirror Symmetry:  Lessons from the $(2,2)$ Locus}
The $(2,2)$ locus of the quintic possesses the wonderful property of mirror symmetry.  Although mirror symmetry is a phenomenon associated
to the equivalence of two conformal theories, in most known examples dual pairs may be constructed in terms of dual pairs of massive
linear models \cite{Morrison:1995yh}.  For example, in the case of the quintic we can write down a mirror linear model with $101$ 
complexified K\"ahler parameters and a single complex structure modulus.  The identification of the parameters between the theories is 
complicated by the usual mirror map\footnote{In the linear model this is generated by the point-like instantons in the gauge theory 
\cite{Witten:1993yc}.} but up to this coordinate re-parametrization, we can easily identify the mirror theories.  

As we discussed in the introduction, there are good reasons to believe that there should be a generalization of mirror symmetry off the
$(2,2)$ locus.  Assuming this generalization exists, what may we learn about it from the form of the Lagrangian? The simplest guess would
be that given a $(0,2)$ CFT arising as the IR limit of a linear model, there ought to exist a linear model that flows to the dual theory.  Thus,
a linear model specified by the data $\tau^a, E^i, J_i$, should have a dual specified by $\tauh^A, \Eh^I, \Jh_I$.\footnote{To be precise about 
this exchange, we would need to develop analogues of ``toric'' K\"ahler deformations and ``polynomial'' complex structure/bundle deformations
\cite{Aspinwall:1993nu}.  These are the deformations of the CFT that may be easily identified with UV parameters of the linear model.}  On the 
$(2,2)$ locus we can split the $J_i$ deformations into those that preserve $(2,2)$ SUSY---$J_i^{(2,2)}$
and those that only preserve $(0,2)$ SUSY---$J_i^{(0,2)}$.  The usual $(2,2)$ mirror symmetry suggests that {\em on the $(2,2)$ locus} we can
identify the complex structure parameters with the mirror K\"ahler parameters, and vice versa:
\begin{eqnarray}
\tau^a	 	    & \leftrightarrow & \Jh_I^{(2,2)}, \nonumber\\
J_i^{(2,2)}             & \leftrightarrow & \tauh^A,
\end{eqnarray}
and at least infinitesimally, we would expect to identify the $(0,2)$ deformations as
\begin{eqnarray}
J_i^{(0,2)}  &\leftrightarrow & \Eh^I, \nonumber\\
E^i		 &\leftrightarrow & \Jh_I^{(0,2)}.
\end{eqnarray}

\subsection{$(0,2)$ Mirror Symmetry for Fano Varieties}
The full set of these deformations will be difficult to study all at once, and an important simplification
may be made by setting $J_i = 0$ and choosing the combinatoric data so that the moduli space of the
linear model is compact and toric.  Although these theories are gapped, their chiral ring contains 
holomorphic data generalizing the familiar Gromov-Witten 
invariants to include bundle deformations. 

On the $(2,2)$ locus it is known that there is a notion of mirror symmetry even for these massive
theories.  The basic observation is that one may use a dual Landau-Ginzburg description to compute the chiral ring of these
theories.  This description is easy to obtain either by using a dualization procedure as in~\cite{Hori:2000kt}, or by doing computations on the 
Coulomb branch of the linear model~\cite{Melnikov:2006kb}.

The work of~\cite{Adams:2003zy}\ suggests that this dual description should also exist off the $(2,2)$ locus. 
Based on that work, it appears that in this class of models the generalized mirror map
should relate the $\tau,E$ deformations of the GLSM to superpotential (i.e. $J_i$) deformations of a dual LG
model.  To test this, it is important to learn to compute correlators in $(0,2)$ Landau-Ginzburg theories.
This is the issue to which we turn in the next section.

\section{Half-Twisted Landau-Ginzburg Models}\label{s:lg}
To construct the Landau-Ginzburg theories of interest, we can simply drop most of the terms in the
linear model Lagrangian.  Dropping the gauge fields, we are left
with a simple action: $S = S_{\text{kin}}+ S_J$:
\begin{eqnarray}
\label{eq:LGaction}
S_{\text{kin}} & = & \int d^2 y d^2\theta \left\{ - \ff{i}{2} \Phib^i \p_- \Phi^i - \ff{1}{2} \Gammab^i \Gamma^i\right\}, \nonumber\\
S_J & = & -\ff{1}{\sqrt{2}} \int d^2 y d\theta^+ \Gamma^i J_{i} (\Phi)|_{\thetab^+ =0} + \text{h.c.}.
\end{eqnarray}
The component Lagrangian in Euclidean signature is given by 
\begin{eqnarray}
\cL &=& 4 \p_{z} \phi^i \p_{\zb} \phib^i  + 2\psib_{+}^i \p_z \psi_+^i + 2\gammab_{-}^i  \p_{\zb} \gamma_-^i + J_i \Jb_i 
- J_{i,j}\psi_+^j \gamma_-^i - \Jb_{i,j} \gammab_-^i \psib_+^j\nonumber\\
~&~&+E^i\Eb^i - E^i_{~,j} \psi_+^j \gammab_-^i  - \Eb^i_{~,j} \gamma_-^i \psib_+^j.
\end{eqnarray}
We observe that an identical theory is obtained by switching $\gamma_-$ with $\gammab_-$, and the $J_i$ with the $E^i$.
The $(2,2)$ locus corresponds to $E^i =0$ and $J_{i,j} = J_{j,i}$,  in which case we may write $J_i = \p W/ \p \Phi^i$.

\subsection{The Half-Twist}
The $(2,2)$ theory possesses an axial $R$-symmetry $\GU(1)_B$, with the action 
\begin{equation*}
\begin{array}{|c|c|c|c|c|c|c|}\hline
\ast			& \theta^+		& \phi	& \gamma_-	& \psi_+ \\ \hline
\GU(1)_B		&+1			&0		&+1			&-1	      \\ \hline
\end{array}
\end{equation*}
This R-symmetry may be used to twist the $(2,2)$ theory to obtain the B-twisted topological field theory. Not all $(0,2)$ deformations
preserve $\GU(1)_B$:  while any $J_i$ deformation is allowed, the $E^i$ must remain zero. Thus, the half-twisted theory will only be
of use for the study of the $J_i$ deformations.  Our discussion of the $(0,2)$ linear models and their LG ``mirrors'' suggests that
this restriction is quite natural, at least for the case of linear models without superpotential, and we will now restrict attention to 
LG models with $E^i=0$.

With this simplification, the action of the $(0,2)$ supercharges $Q_+,\Qb_+$ is:
\begin{equation*}
\begin{array}{|c|c|c|c|c|c|c|c|c|}\hline
\ast					& \phi^i		& \phib^i			& \psi_+^i 			& \psib_+^i	& \gamma_-^i 	& \gammab_-^i \\ \hline
\ff{1}{\sqrt{2}}\ACO{Q_+}{\cdot}	&\psi_+^i		&~0				&0				&-2\p_{\zb}\phib^i	&-\Jb_{i}		& 0			\\ \hline
\ff{1}{\sqrt{2}}\ACO{\Qb_+}{\cdot}	&0			&~\psib_+^i		&-2\p_{\zb}\phi^i	&0			&0			& J_{i}		\\ \hline
\end{array}
\end{equation*}

We can now follow the standard procedure for constructing the twisted theory~\cite{Witten:1991zz}. Let $T$ be the Lorentz
generator in the untwisted theory.  The Lorentz generator in the twisted theory is then taken to be $T' = T - \ff{1}{2} V$: 
\begin{equation*}
\begin{array}{|c|c|c|c|c|c|c|c|c|c|}\hline
\ast		& \phi^i		& \phib^i		& \psi_+^i 			& \psib_+^i	& \gamma_-^i 	& \gammab_-^i 		& \Qb_+		\\ \hline
T		&0			&	0		&	\ff{1}{2}		& \ff{1}{2}		&-\ff{1}{2}		&-\ff{1}{2}			&+\ff{1}{2}		\\ \hline
V		&0			&	0		&	-1			& +1			&+1			&-1 				& +1			\\ \hline
T'		&0			&	0		&	+1			& 0			&-1			&0 				& 0			\\ \hline
\end{array}
\end{equation*}
In particular, we see that $\ff{1}{\sqrt{2}}\Qb_+$ becomes a world-sheet scalar fermionic operator, which we will denote by $\Qb$.   
It will also be useful to rename the fermions: 
\begin{eqnarray}
\psi_+ 		& \to &	\rho_{\zb}, \nonumber\\ 
\gamma_- 	& \to &	\eta_z, \nonumber\\
\psib_+ 	    	& \to &	\theta, \nonumber\\
\gammab_- 	& \to	&	\chi.
\end{eqnarray}
 With
this notation, the Lagrangian for the twisted theory takes the form
\begin{eqnarray}
\cL & = & 4 \p_{z} \phi^i \p_{\zb} \phib^i  + 2\rho_{\zb}^i \p_z \theta^i + 2\eta_{z}^i  \p_{\zb} \chi^i + J_i \Jb_i 
- J_{i,j}\rho_{\zb}^j \eta_{z}^i - \Jb_{i,j} \chi^i \theta^j \nonumber\\
~&=& 2 \eta^i_{z} \p_{\zb} \chi^i - J_{i,j} \rho^j_{\zb} \eta^i_z+ \ACO{\Qb}{V},
\end{eqnarray}
where $V = -2 \rho^i_{\zb} \p_z \phib^i + \chi^i \Jb_i$.

So far, we have formulated the theory on a flat world-sheet.  It is not hard to re-write the twisted action in the background of a fixed 
world-sheet metric $g$. We find that the action takes the form
\begin{eqnarray}
S & = & \int_\Sigma \left\{ \ast_g\left[ \phib^i \nabla_d \phi^i + J_i \Jb_i - \chi^i \Jb_{i,j} \theta^j \right]  + \rho^i\wedge\ast_g \pb \theta^i
\right.\nonumber\\
~&~&\left.~~~~~~+i \eta^i \wedge \pb \chi^i + \ff{i}{2} J_{i,j} \eta^i \wedge \rho^j\right\}.
\end{eqnarray}
The $\ast_g$ denotes the Hodge star map constructed with the metric $g$.
The first line above is $\Qb$-exact, so that we may write $S= S' + \ACO{\Qb}{V}$, with
\begin{eqnarray}
S' & = & \int_\Sigma \left\{i \eta^i \wedge \pb\chi^i + \ff{i}{2} J_{i,j} \eta^i\wedge \rho^j \right\},\nonumber\\
V & = & \int_\Sigma \left\{-\rho^i \wedge \ast_g\pb\phib^i + \ast_g\chi^i \Jb_i \right\} .
\end{eqnarray}
Having constructed the Lagrangian, we may now define the half-twisted theory by projecting onto the cohomology of $\Qb$.
That is, we compute correlators of $\Qb$-closed operators. These decouple from $\Qb$-exact terms, in the sense that
\begin{equation}
\la \O_1(x_1) \cdots \O_k (x_k) \ACO{\Qb}{W} \ra = 0
\end{equation}
for all $\Qb$-closed $\O_i$ and all $W$.  In what follows, we will be concerned with correlators of $\Qb$-closed local operators.  
Inspection of the $\Qb$ action shows that these are generated by $\phi^i(z,\zb)$ modulo the relations $J_i(\phi) = 0$.  

\subsection{Twisting versus Half-Twisting}
The half-twisted theory superficially resembles the familiar twisted topological theory obtained from a $(2,2)$ Landau-Ginzburg theory.
There, in addition to $\Qb_+$, $\Qb_-$ also becomes a world-sheet scalar under the $\GU(1)_B$ twist, and one may use
$Q = \Qb_+ + \Qb_-$ as the BRST operator.  In this case, the twisted action may be written as $$S = S_{\text{top}}  + \ACO{Q}{V},$$
with $S_{\text{top}}$ a term independent of the world-sheet metric.  This result has two important consequences.  First, the 
energy-momentum tensor is $Q$-exact, and it follows that the correlators of local $Q$-closed operators are independent of 
the positions of the insertions.  Second, we may use the metric independence of the twisted theory to simplify computations:
on the one hand, rescaling the world-sheet metric $g \to \lambda g$ is a $Q$-trivial deformation of the action, while on the other
hand as $\lambda \to \infty$, the semi-classical expansion about the critical points of the potential becomes an arbitrarily good
approximation \cite{Vafa:1990mu}.  In the $(2,2)$ case, this leads to a simple formula for correlators of local observables 
on a Riemann surface of genus $h$:
\begin{equation}
\label{eq:22LG}
\la \phi^{i_1}(x_1) \cdots \phi^{i_k} (x_k) \ra_h = \sum_{\phi_\ast} \phi^{i_1}_\ast \cdots \phi^{i_k}_\ast \left[\det \Hess W\right]^{h-1},
\end{equation}
where $\phi_\ast$ denotes a critical point of the superpotential.

Do these results survive off the $(2,2)$ locus? Examination of the $(0,2)$ action suggests a negative answer, 
since $S'$ depends on the complex structure chosen on the world-sheet.  A choice of complex structure is equivalent
to the choice of projectors on the space of $1$-forms on the Riemann surface: $\pi: \Omega^1 \to \Omega^{(1,0)}$ and 
$\pit: \Omega^1 \to \Omega^{(0,1)}$.  Combining $\rho$ and $\eta$ into a single fermionic 1-form $\kappa = \eta_z dz+\rho_{\zb} d\zb$,
we have
\begin{equation}
S' = \int_\Sigma \left\{i \pi \kappa^i \wedge \pit d \chi^i + \ff{i}{2} J_{i,j} \pi\kappa^i \wedge \pit \kappa^j\right\}.
\end{equation}
Thus variations of the complex structure modify the action.
A closely related problem is that the energy-momentum tensor is no longer exact, and seemingly we can no longer argue 
that the correlators are position-independent.   Instead, by following the $(2,2)$ argument, we will find that the correlators are
only constrained to depend holomorphically on the positions of the insertions; see, for example~\cite{Witten:2005px}.  

Fortunately, there is one feature that the twisted and half-twisted theories do have in common:  a constant rescaling of the
world-sheet metric modifies the action by BRST-trivial terms.  Thus, even in the $(0,2)$ case, we may compute the
correlators exactly by working in the limit of an arbitrarily large Riemann surface.  In this limit fluctuations about the vacua are 
suppressed, and we expect the semi-classical expansion to become exact.  Furthermore, in {\em gapped} $(0,2)$ theories it is 
clear that in this limit correlators become independent of positions of the inserted operators.\footnote{We expect this to be true in the conformal case as well but  this calls for a more careful analysis.}
Since this rescaling of the metric is $\Qb$-exact, 
we conclude that the correlators of local $\Qb$-closed operators will be position-independent and are computed 
exactly by the semi-classical expansion about the vacua.   

The spirit of this discussion should be familiar from the work on generalizing the structure of the chiral rings to $(0,2)$ 
theories \cite{Adams:2003zy, Adams:2005tc}.  These authors faced similar issues in discussing gapped non-linear sigma models, and they
relied on $\Qb$-triviality of constant rescalings to argue for the existence of the ring structures off the $(2,2)$ locus.  It is not surprising
that we should be able to use similar arguments in the simpler setting of Landau-Ginzburg theories.  Indeed, here the arguments
are more transparent because all the problems and their resolutions are already visible at tree-level, while in the non-linear models 
one has to carefully consider the one-loop beta function and the corresponding fact that quantum-mechanically 
$T_{z\zb}$ fails to be $\Qb$-exact.

Our basic conclusion is that, at least as far as genus zero correlators of local operators are concerned, the 
half-twisted gapped  Landau-Ginzburg theory is not sensitive as to whether it is on or off the $(2,2)$ locus.  The 
dependence on the world-sheet complex structure is an interesting feature of the theory on higher genus Riemann 
surfaces.  It is not clear to us whether it will show up in correlators of local observables, or will only be sensitive to insertions of non-local
operators.  Since to discuss higher genus amplitudes we would need to couple the theory to world-sheet gravity, we will
not discuss this depdendence further.  {\em In what follows, we will restrict attention to genus zero correlators.}  

The form of the correlators on the $(2,2)$ locus (eqn.(\ref{eq:22LG})) leads to an obvious guess for the form of the correlators in these LG models:
\begin{equation}
\label{eq:corrs}
\la \phi^{i_1}(x_1) \cdots \phi^{i_k} (x_k) \ra = \sum_{\phi_\ast} \phi^{i_1}_\ast \cdots \phi^{i_k}_\ast\left[\det_{i,j} J_{i,j}\right]^{-1},
\end{equation}
where the sum is over the classical vacua, $J_i(\phi_\ast) = 0$.  In the next section we will derive this expression by
a simple semi-classical expansion.
 
\section{The Semi-Classical Expansion} \label{s:semiclass}

\subsection{The Free Theory}
We begin by solving the free theory on the sphere. We take $J_i =  m_{ij} \phi^j$.  The action is quadratic in the fields, so we should 
have no trouble computing all the correlation functions.  The computation carried out here is entirely straightforward, and as expected,
we will find that a non-zero answer is entirely due to the zero modes.  The non-zero modes cancel out as a consequence
of supersymmetry.  We emphasize that this is to be expected, but we nevertheless present the computation because it is
the simplest hands-on example, and it certainly guides our intuition for these sorts of $(0,2)$ determinants.

It will be convenient to deal separately with the bosons and fermions, so we will split the
action accordingly:  $S= S_{b} + S_{f}$:
\begin{eqnarray}
S_{b} & = & \int_\Sigma \phib^i \left[ \Delta_d \delta^{ij} + \bar{m}_{ki} m_{kj} \right] \phi^j, \nonumber\\
S_{f}  & = & \int_\Sigma \rho^i \wedge \ast_g \pb \theta^i + i \eta^i \wedge \pb \chi^i + \ff{i}{2} m_{ij} \eta^i\wedge \rho^j - \ast_g\bar{m}_{ij} \chi^i \theta^j.
\end{eqnarray}

We will solve the theory by a mode expansion in terms of the eigenfunctions of the Laplacian on the sphere:
\begin{equation}
\Delta_d f_\alpha = \lambda_\alpha^2 f_\alpha,~~~ \int_\Sigma f_\alpha\wedge \ast_g f_\beta  = \delta_{\alpha\beta}.
\end{equation}
Since the Riemann surface is K\"ahler, it follows that  $||\p f_\alpha||^2 = \lambda_\alpha^2/2.$
We may use these eigenfunctions to expand all the fields.  To keep notation simple we will drop 
the flavor indices.  We write the expansion as 
\begin{eqnarray}
\phi    & = & \phi_0/\sqrt{\Vol_g}    +  \sum_\alpha \phi_\alpha f_\alpha, \nonumber\\
\theta & = & \theta_0/\sqrt{\Vol_g} +  \sum_\alpha \theta_\alpha f_\alpha, \nonumber\\
\chi & = & \chi_0/\sqrt{\Vol_g} +  \sum_\alpha \chi_\alpha f_\alpha, \nonumber\\
\eta & = & \sum_\alpha \ff{2}{\lambda_\alpha} \eta_\alpha \p f_\alpha ,\nonumber\\
\rho & = & \sum_\alpha \ff{2}{\lambda_\alpha} \rho_\alpha \pb f_\alpha .
\end{eqnarray}
The only change in this expansion for genus greater than zero would be the introduction of $\eta$ and $\rho$ zero 
modes. It is these that may yield a dependence on the world-sheet complex structure.  

The action splits into a sum over the modes: $S = S^0 + \sum_\alpha S^\alpha$, so that we can perform the integration
mode by mode.  Let us first treat the zero modes.   The action is
\begin{eqnarray}
S_{b}^0 & = & \phib^i_0 ~ \bar{m}_{ki} m_{kj} ~\phi_0^j, \nonumber\\
S_{f}^0  & = & -\chi^i_0 \bar{m}_{ij} \theta_0^j .
\end{eqnarray}

Unlike a generic $(0,2)$ theory, the models with a $(2,2)$ locus possess a canonical choice for
the path integral measure, and with this choice we have
\begin{equation}
D[\text{fields}_0] = \prod_{i}  \frac{d^2 \phi_0^i}{\pi} d\chi^i_0 d\theta^i_0 
\end{equation}
for the zero modes.  This leads to
\begin{equation}
D[\text{fields}_0] e^{-S^0} = \frac{1}{\det m^\dag m} \det m^\dag = (\det m)^{-1}.
\end{equation}

The non-zero modes are not much harder.  We can write the action as
\begin{eqnarray}
S_{b}^\alpha & = &=\phib^i_\alpha \left[ \lambda^2_\alpha \delta_{ij} + \left(m^\dag m\right)_{ij} \right] \phi^j_\alpha,\nonumber\\
S_f^\alpha &=& \rho^i_\alpha \lambda_\alpha \delta_{ij} \theta^j_\alpha -\chi^i_\alpha \lambda_\alpha \delta_{ij} \eta^j_\alpha
                     -\rho^i_\alpha \left(^t m\right)_{ij} \eta^j_\alpha - \chi^i_\alpha \bar{m}_{ij} \theta^j_\alpha.
\end{eqnarray}
It is convenient to re-write the fermion action by combining $\rho^i_\alpha, \chi^i_\alpha$ into a row vector
$\Psib_\alpha$, and $\eta_\alpha^i, \theta_\alpha^i$ into a column vector $\Psi_\alpha$.  In terms of these variables,
the fermion action takes the form
\begin{equation}
S_f^\alpha =- \Psib_\alpha^I A_{IJ}^\alpha \Psi^J_\alpha,
\end{equation}
with $A$ given by
\begin{equation}
A^\alpha = 
\left( \begin{array}{cc}		-\lambda_\alpha \mathbbm{1}	& 	^t m		\\
						\bar{m}			&	+\lambda_\alpha \mathbbm{1}
	\end{array}
\right).
\end{equation}
The measure for the non-zero modes is again canonical:
\begin{equation}
D[\text{fields}_\alpha] = \int \prod_i \frac{d^2\phi^i_\alpha}{\pi} d \eta^i_\alpha d\rho^i_\alpha d\theta^i_\alpha d\chi^i_\alpha.
\end{equation}
The bosonic integral yields $\det\left[\lambda_\alpha^2+m^\dag m\right]^{-1}$, and the fermionic integration gives
$\det (-A^\alpha)$, which is easily verified to be the inverse of the bosonic contribution.   So, indeed, as expected,
the non-zero mode contributions cancel model by mode. Thus, it is easy to see that, at least in the free theory, regularization
is not an issue.

Finally, we can write down the correlators in this trivial example.  The only non-trivial observable is the identity, and
we conclude that
\begin{equation}
\la 1 \ra = \left(\det m\right)^{-1}.
\end{equation}

\subsection{Correlators via Localization}
Having disposed of the free field case, we are ready to tackle the more interesting theory with arbitrary $J_i$.  Because
we are projecting onto the $\Qb$ cohomology, we expect that we should be able to do all computations by a semi-classical
computation about $\Qb$-invariant field configurations.  These are also the configurations that dominate the path
integral when we scale up the metric on the Riemann surface.  Examining the $\Qb$ variations, we conclude that the
field configurations are $\phi = \phi_\ast$ satisfying  $\pb \phi_\ast^i = 0$ and $J_i(\phi_\ast) = 0$.  In the case of a gapped
theory, the latter in fact requires $\phi_\ast$ to be constant on the world-sheet, so that the $\Qb$-invariant configurations
consist of a finite number of points---the simultaneous zeroes of $J_i = 0$.  Thus, the correlators are given by
\begin{equation}
\la \phi^{i_1}(x_1) \cdots \phi^{i_k}(x_k) \ra = \sum_{\phi_\ast} \la  \phi^{i_1}(x_1) \cdots \phi^{i_k}(x_k) \ra^\text{free} |_{m_{ij} = J_{i,j}(\phi_\ast)},
\end{equation}
with $\phi^i = \phi^i_\ast + \phi'^i$, where $\phi'i$ are the bosonic fields in the free theory.  Using our solution of the free field
correlators, we find the expression advertised in eqn. (\ref{eq:corrs}).

\subsection{Properties of the Correlators}
We will now use the general expression to derive some properties satisfied by the correlators.  First, since the
 correlator is obtained by summing over the zeroes of $J_i = 0$, it is clear that our correlators obey the ``quantum 
 cohomology relations''
\begin{equation}
\la \O J_i \ra = 0 ~~~\text{for all} ~\O~.
\end{equation}

Second, it is easy to derive selection rules that lead to constraints on the correlators.   Consider a
global  $\GU(1)$ symmetry of the free theory, where the $\Phi^i$ multiplets have charges $q_i$, and the 
$\Gamma^i$ have charges $-Q_i$.  Suppose the $J_i$ depend on parameters $t_A$.  We can always assign these 
parameters charges $Q_A$ such that $J_i$ has charge $Q_i$.  In terms of the twisted fields, we have

\begin{equation*}
\begin{array}{|c|c|c|c|c|c|c|c|c|c|c|}\hline
\ast		& \phi^i		& \phib^i		& \rho^i 	& \eta^i	& \theta^i 		& \chi_-^i 		& t_A	& J_i		\\ \hline
\GU(1)	& q_i			& -q_i		& q_i 	& -Q_i	& -q_i	 	& Q_i 		& Q_A	&Q_i	\\ \hline
\end{array}
\end{equation*}
\noindent
as a symmetry of the action.  When $Q_i \neq q_i$, this is an anomalous symmetry, but of
course it may still be used to derive selection rules.  The transformation of the measure is easy to see in the
twisted variables, where it is just due to the zero-mode mismatch.  Under a field redefinition 
$\rho^i \to e^{i\alpha q_i}\rho^i$, $\eta^i \to e^{-i Q_i \alpha}\eta^i$ and so on, we find that the measure shifts by
$D[\text{fields}] \to D[\text{fields}] \Delta$, where
\begin{equation}
\Delta(\alpha) = \prod_{i=1}^N e^{i\alpha(q_i-Q_i)}.
\end{equation}
This leads to the selection rule
\begin{equation}
\la \phi^{i_1}\cdots \phi^{i_k}\ra (e^{i\alpha Q_A}t_A) = \Delta(\alpha) \exp\left(i\alpha\left[q_{i_1}+\cdots+q_{i_k}\right]\right\}
\la \phi^{i_1}\cdots \phi^{i_k}\ra(t).
\end{equation}
As usual in supersymmetric theories, these rules are quite powerful once they are combined with constraints from holomorphy.

Finally, we note that just as in the case of the $(2,2)$ correlators, the $(0,2)$ correlators may be expressed by a residue
formula:
\begin{equation}
\la \phi^{i_1} \cdots \phi^{i_k} \ra = \frac{1}{(2\pi i)^N}\int_{C} \frac{d\phi^1 \wedge d\phi^2 \wedge\cdots\wedge d\phi^N}{J_1 J_2\cdots J_N} \phi^{i_1} \cdots \phi^{i_k},
\end{equation}
where $C$ is a multicontour for the set of common zeroes of the $J_i$\cite{Vafa:1990mu,Griffiths:5}. 


\section{Examples} \label{s:examples}
\subsection{The $(2,2)$ Example}
The simplest (2,2) LG theory that has a $(0,2)$ deformation is given by a massive deformation of the product of
$A_k$ and $A_{p}$ $N=2$ minimal models.  On the $(2,2)$ locus, we have the superpotential
\begin{equation}
W = \ff{1}{k+1} X^{k+1} - t X + \ff{1}{p+1} Y^{p+1} - s Y.
\end{equation}
In $(0,2)$ language, we have
\begin{eqnarray}
J_X &=& X^k - t \nonumber\\
J_Y &=& Y^p -s.
\end{eqnarray}
This simple theory has a discrete R-symmetry $\Z_k \times \Z_p$, under which $X$ has charges $(1/k,0)$ and $Y$ has
charges $(0,1/p)$.  Recalling the Jacobian for the measure, we see that this symmetry implies that the 
$\la X^A Y^B \ra$ vanish unless $A=k-1\mod k$ and $B=p-1 \mod p$. The quantum cohomology relations reduce 
the non-vanishing correlators to
\begin{equation}
\la X^{mk+k-1} Y^{np +p-1} \ra = t^m s^n \la X^{k-1} Y^{p-1} \ra.
\end{equation}
Finally, using our formula, we see that 
\begin{equation}
\la X^{k-1} Y^{p-1} \ra = \sum_{(X_\ast,Y_\ast)} \frac {X_\ast^{k-1} Y_\ast^{p-1}}{kp X_\ast^{k-1} Y_\ast^{p-1}} = 1.
\end{equation}

\subsection{A $(0,2)$ deformation}
The simple theory we just solved has plenty of $(0,2)$ deformations.  As a concrete example, let us pick 
\begin{eqnarray}
J_X &=& X^k - t - \alpha Y, \nonumber\\
J_Y &=& Y^p -s.
\end{eqnarray}
The $(2,2)$ locus is given by $\alpha = 0$.  The quantum cohomology relations yield the constraints
\begin{eqnarray}
\la X^A Y^B\ra  & = & t \la X^{A-k} Y^B \ra +\alpha \la X^{A-k} Y^{B+1}\ra, \nonumber\\
\la X^A Y^B \ra & = & s \la X^A Y^{B-p} \ra.
\end{eqnarray}
Thus, the undetermined correlators have $A < k$ and $B <p$.  Furthermore, the theory still has a $\Z_k$ symmetry
under which $X$ has charge $1/k$ and $Y$ is neutral.  Applying the same considerations as above, we see that we
just need to consider the correlators with $A=k-1$. The quickest way to proceed is again to just use the explicit form for the correlators:
\begin{equation}
\la X^{k-1} Y^{B} \ra = \sum_{(X_\ast,Y_\ast)} \frac {X_\ast^{k-1} Y_\ast^{B}}{kp X_\ast^{k-1} Y_\ast^{p-1}} = \frac{1}{p}\sum_{Y_\ast} Y^{B-p+1}.
\end{equation}
Thus, the non-vanishing correlators have $B=np+ p-1$, and  $\la X^{k-1} Y^{p-1} \ra = 1.$
Using this, it is easy to see that there are new non-vanishing correlators when $\alpha \neq 0$.  For example, we find
\begin{equation}
\la X^{mk +k-1} Y^{p-1-m} \ra = \alpha^m.
\end{equation}

We can use this example to illustrate a simple trick often useful in these computations.  This is essentially a specialization of
the more general residue results mentioned above.  To compute $\la X^A Y^B\ra$ we need to
sum over the simultaneous zeroes of $J_X $ and $J_Y$.  We can solve for $X^k$ in terms of $Y$, and then be left with a sum over
the roots of $Y^p =s$.  Using the the familiar relation
\begin{equation}
\sum_{z_\ast|P(z_\ast)=0} f(z_\ast) = \oint_C \frac{dz}{2\pi i} f(z) \frac{P'(z)}{P(z)},
\end{equation}
where $C$ encloses the roots of the polynomial $P(z)$, we are left with an elegant formula for the correlators:
\begin{equation}
\la X^{mk+k-1} Y^B\ra = \oint_C \frac{dz}{2\pi i} \frac{ (t+\alpha z)^m z^B}{z^p -s},
\end{equation}
where $C$ is a contour given by $|z| = |s|+\ep$ for some positive $\ep$.  In this case, it is a simple matter to pull the contour
around to encircle $z = \infty$, where the residue is easy to evaluate, leading to the non-zero correlators being given by
\begin{equation}
\la X^{mk+k-1} Y^{K+p-1-m}\ra = \alpha^m \left(\frac{t}{\alpha}\right)^K \left.\frac{1}{K!}\frac{d^K}{du^K} \frac{(1+u)^m}{1-s \frac{\alpha^p u^p}{t^p}}\right|_{u=0}.
\end{equation}
More generally, this kind of approach makes it clear that there is never any need to evaluate specific roots---a good thing, as
this would make the problem quite nasty.  In fact, all of our computations may be performed algebraically.

\subsection{The $(0,2)$ mirror of $\P^1\times \P^1$}

A generalization of the Hori-Vafa dualization procedure for $(0,2)$ models was proposed in~\cite{Adams:2003zy}.
The starting point for the most interesting example in~\cite{Adams:2003zy}\ is the $(2,2)$ GLSM for $\P^1\times \P^1$.   We will
now describe the linear model, the dual Landau-Ginzburg theory proposed in \cite{Adams:2003zy}, and we will then  compute 
correlators in the dual theory. 

The field content of this theory is easy to get by adapting the action for the quintic given above.  To get a $\P^1$ model, we need two
chiral fields $\Phi^i$ and their Fermi multiplets $\Gamma^i$, $i=1,2$ that are minimally coupled with charge $1$ to a 
gauge field $V$ with field-strength $\Upsilon$. To be on the $(2,2)$ locus, we also need to introduce a neutral chiral
field $\Sigma$.  Gauge invariance forbids non-zero $J_i$,  and $(2,2)$ SUSY requires $E^i = i \sqrt{2} \Sigma \Phi^i$.
The theory depends on a single complexified F-I parameter $\tau$ corresponding to the complexified K\"ahler class
of the $\P^1$. 

To get the theory considered in~\cite{Adams:2003zy}, we introduce a second $\P^1$, with field content $\Phit^i,\Gammat^i$, $\Vt$,  $\Upsilont$,
$\Sigmat$, and another F-I parameter $\taut$. This $\P^1\times \P^1$ theory has $(0,2)$ deformations that are visible in the
linear model.  Following \cite{Adams:2003zy}, we consider
\begin{eqnarray}
E^1     & = & i\sqrt{2} \left\{ \Phi^1 \Sigma + \Sigmat (\alpha_1 \Phi^1 + \alpha_2 \Phi^2)\right\},\nonumber\\
E^2     & = & i\sqrt{2} \left\{ \Phi^2 \Sigma + \Sigmat (\alpha_1' \Phi^1 + \alpha_2' \Phi^2)\right\},\nonumber\\
\Et^1   & = & i\sqrt{2} \left\{ \Phit^1 \Sigmat + \Sigma (\beta_1 \Phit^1 + \beta_2 \Phit^2)\right\},\nonumber\\
\Et^2   & = & i\sqrt{2} \left\{ \Phit^2 \Sigmat + \Sigma (\beta_1' \Phit^1 + \beta_2' \Phit^2)\right\}.
\end{eqnarray}
Not all of these deformation parameters are independent:  sending $(\alpha,\alpha') \to (\lambda\alpha,\lambda\alpha')$ and
$(\beta,\beta') \to (\mu \beta,\mu\beta')$ corresponds to trivial deformations for any non-zero $\mu$,$\lambda$.  Thus, there
are a total of six independent $(0,2)$ deformations.

Dualizing this model leads 
to a $(0,2)$ Landau-Ginzburg theory with two chiral superfields $X,Y$ and
\begin{eqnarray}
\label{eq:ABSJ}
J_X & = & X +\frac{P_1}{X} + q_1 Y +  \frac{S_1}{Y}, \nonumber\\
J_Y & = & Y + \frac{P_2}{Y} + q_2 X + \frac{S_2}{X},
\end{eqnarray}
where $q_1,q_2$ are some functions of $\alpha,\beta$, and 
\begin{equation}
\begin{array}{ccccccc}
P_1		& = &-(1+p_1(\alpha,\beta))e^{2\pi i \tau},		& ~~~  &  S_1 & = & -s_1(\alpha,\beta) e^{2\pi i\taut}, 	\\
P_2		& = &-(1+p_2(\alpha,\beta))e^{2\pi i \taut},		& ~~~  &  S_2 & = & -s_2(\alpha,\beta) e^{2\pi i\tau} .
\end{array}
\end{equation}
The $(2,2)$ locus corresponds to $p_i=s_i=q_i = 0$.

Like the Landau-Ginzburg theories we have been discussing, the $(0,2)$ linear model admits a half-twist, and one expects
that in a massive model of the sort just described, one should still be able to compute correlators of half-chiral observables.
In this case, it would amount to computing correlators of the form $\la \sigma^a \sigmat^b\ra$, where the $(\sigma,\sigmat)$ are the
lowest components of the $(\Sigma,\Sigmat)$ multiplets.  Recently, Guffin and Katz \cite{Guffin:2007mp} computed a number of correlators 
in this half-twisted linear model, expanding upon earlier work of Katz and Sharpe\cite{Katz:2004nn}.  By studying the moduli space of 
instantons of degrees $0,1,2$, Guffin and Katz were able to compute a number of correlators for a generic $(0,2)$ deformation of the theory.  
This required an impressive array of algebraic techniques and substantial computational effort, and it yielded a quite elegant form for
a number of  correlators.

Our simple observations on the $(0,2)$ Landau-Ginzburg correlators now allow us to compare the results of Adams et.~al.~with those 
of Guffin and Katz.  By computing the correlators explicitly, we will be able to make a quantitative test of generalized $(0,2)$ 
mirror symmetry.  At this time the story is still incomplete and, as has been discussed in \cite{Guffin:2007mp}, the real issue is the form of the mirror map.  In fact, there are, in principle, three separate issues:
\begin{itemize}
\item What is the precise form of the $(p, q, s)$ as functions of $(\alpha, \beta, \alpha', \beta')$?
\item Is there a parameter-dependence in the relation between insertions of $\sigma,\sigmat$ and insertions of $X,Y$?  We know this is not the
         case on the $(2,2)$ locus, but it is possible (and likely) that this is modified for a generic $(0,2)$ deformation.
\item The dualization produces an additional factor in the measure of the theory.  On the $(2,2)$ locus, this
         amounts to replacing $\det J$ with $X Y \det J$.  Is this measure factor modified off the $(0,2)$ locus?
\end{itemize}
We will not address these issues in this paper.  Instead, we will content ourselves with a solution of the model 
of~\cite{Adams:2003zy}, and leave the applications of this solution to upcoming work \cite{Guffin:unpub}.

With this modest goal in mind, we return to the $J_X,J_Y$ of eqn.~(\ref{eq:ABSJ}).  Elementary counting of solutions reveals
that generically, the zeroes consist of four points in $(\C^\ast)^2$.  Although at first the equations look a little formidable, a
simple trick renders them quite simple.  We can ``integrate in'' a field $Z$ with $J_Z = Z - XY$ without changing the chiral 
ring.  Working in this theory, and first solving for $(X,Y)$, we find that the vacua satisfy
\begin{eqnarray}
P(Z) & = & (1-q_1 q_2) Z^2 + (S_1+S_2 -q_1 P_2 -P_1 q_2) Z + S_1 S_2 -P_1 P_2 = 0, \nonumber\\
X 	& = & -\frac{q_1 Z +P_1}{Z +S_1} Y, ~~~
Y^2 	 =  - \frac{Z (Z +S_1)}{q_1 Z +P_1}.
\end{eqnarray}
The correlators are now given by the sum 
\begin{equation}
\la X^A Y^B \ra = \sum_{Z_\ast,Y_\ast} X_\ast^A Y_\ast^B \det J^{-1} \left[X_\ast Y_\ast\right]^{-1},
\end{equation}
where the last factor comes about from the change in measure proposed by~\cite{Hori:2000kt}.

With this form, a simple Maple routine readily yields correlators.  For instance, we find
\begin{eqnarray}
\la X X \ra & = & \frac{s_1 -q_1 -q_1 p_2}{(1+p_2-s_1 q_2)(1-q_1 q_2)}, \nonumber\\
\la X  Y\ra & = & \frac{1}{1-q_1 q_2}, \nonumber \\
\la Y Y \ra & = & \frac{s_2 -q_2 -q_2 p_1}{(1+p_1-s_2 q_1)(1-q_1 q_2)}. \nonumber\\
\end{eqnarray}  
Two of the three correlators vanish on the $(2,2)$ locus, but sufficiently far away they reveal interesting
singularity structure.  A naive comparison between these correlators and the results of Guffin and Katz suggest
that the singularity at $q_1 q_2 = 1$ corresponds to a singularity in the linear model correlators, while
the singularities at $s_1 q_2 -p_2 =1$ and $s_2 q_1 -p_1 =1$ may be due to singular terms in the
map from the $\sigma, \sigmat$ to the dual variables $X, Y$.   Further work is necessary to
determine whether this is indeed correct, and we hope to report on these issues soon.

\subsection{A Conformal Example}
As our final example, we will show that it is also possible to apply our results to correlators in conformal theories.  
As a simple example of this,
consider the Landau-Ginzburg theory with superpotential
\begin{equation}
W = \ff{1}{3} (X_1^3+X_2^3+X_3^3) - y X_1 X_2 X_3.
\end{equation}
This theory flows to a CFT. A $\Z_3$ orbifold of this CFT with action 
$$(X_1,X_2,X_3)\to (\zeta X_1,\zeta X_2,\zeta X_3),$$ 
with $\zeta^3=1$, corresponds to the CFT on 
$T^2$ constructed as the hypersurface $W=0$ in $\P^2$.  The easiest way to see this is to construct 
the torus in a linear model and then consider the Landau-Ginzburg phase.  We will compute the unique 
three-point function in the untwisted sector: $C(y) = \la  X_1 X_2 X_3  \ra$.   We expect to find a singularity
in the correlator at $y^3=1$.\footnote{This is of course a little misleading.  This singularity does not correspond
to a real singularity in the torus SCFT.  It is the avatar of the large complex structure limit, and the seeming
singularity may be removed by a proper renormalization of the operators.}

To compute this correlator, we will follow the most straight-forward route:  we will deform the superpotential
with a relevant deformation, compute the correlators, and then take the deformation parameter to zero.  A
simple deformation that does the job is to take:
\begin{equation}
W = \ff{1}{3} (X_1^3+X_2^3+X_3^3) - y X_1 X_2 X_3 - t^2 (X_1+X_2+X_3).
\end{equation}
This superpotential has eight critical points arranged in orbits of the $\Z_3 \subset S_3$ that permutes the $X_i$:
\begin{equation}
\begin{array}{c}
(Y,Y,Y),\\
(Z,Z,-(1+y)Z),\\
(Z,-(1+y)Z,Z),\\
(-(1+y)Z,Z,Z),
\end{array}
\end{equation}
where $Y^2= t^2/(1-y)$ and $Z^2 = t^2/(1+y+y^2)$.  The Hessian is given by
\begin{equation}
H = X_1X_2 X_3 \left[8 (1-y^3) -2y^2 t^2\frac{X_1 + X_2+X_3}{X_1 X_2 X_3}\right].
\end{equation}

There are two ways in which these critical points may degenerate:  when $y^3=1$, we see that all the points run off
to infinity, while when $y=-2$, the critical points collapse to $(\pm\ff{1}{\sqrt{3}},\pm\ff{1}{\sqrt{3}},\pm\ff{1}{\sqrt{3}})$.
While we expect the singularity from $y^3=1$ to survive the $t\to0$ limit, we had better find that troubles at $y=-2$
are illusory.  This is indeed the case: 
\begin{equation}
\begin{array}{ccccccc}
C(y) & = & \left. \frac{X_1 X_2 X_3}{H}\right|_Y  & +  & \left. \frac{X_1 X_2 X_3}{H}\right|_Z & ~ &~\\
~      & = & \frac{2}{2 (1-y)(y+2)^2}			 & +  & \frac{6(1+y)}{2(1-y^3)(y+2)^2}          & = & \frac{1}{1-y^3}.
\end{array}
\end{equation}

Now let us consider $(0,2)$ deformations of this example.  We take
\begin{eqnarray}
J_1 & = & X_1^2 -y_1 X_2 X_3 -t^2, \nonumber\\
J_2 & = & X_2^2 -y_2 X_3 X_1 -t^2, \nonumber\\
J_3 & = & X_3^2 -y_3 X_1 X_2 -t^2.
\end{eqnarray}
We note that these deformations cannot be undone by a field-redefinition. The $(2,2)$ locus is $y_1=y_2=y_3$.  To 
compute $C(y_1,y_2,y_3)$ we can use standard elimination theory to algebraically solve for $X_2$ and $X_3$ in 
terms of $X_1$, leaving us with an eighth-order polynomial for $X_1$.  Alternatively, we may use constraints from 
the $(2,2)$ locus, permutation symmetry, and simple selection rules to constrain the correlator.  Either way, we find 
a simple answer:
\begin{equation}
C(y_1,y_2,y_3) = \frac{1}{1-y_1 y_2 y_3}.
\end{equation}
As in the massive example, we see that the correlator reveals additional singularity structure of the LG theory off the
$(2,2)$ locus.


\section{Conclusions} \label{s:conc}

We have computed correlators of chiral observables in $(0,2)$ deformations of $(2,2)$ Landau-Ginzburg theories and 
have found a simple result:  the correlators are given by a weighted sum over the supersymmetric vacua.  In other
words, in these theories the $(2,2)$ locus is not distinguished {\em even by computability}.  We have  presented
several concrete examples illustrating the general result, as well as a few simple techniques useful in explicit
computations. 

This simple result means that in a large  class of $(0,2)$ theories, important RG-invariants may be computed by simple algebraic 
techniques.  The computation of correlators in conformal LG models follows in an obvious fashion, and we suspect that when
generalized to the gauged linear sigma model, our results may well lead to a $(0,2)$ version of the Morrison-Plesser quantum 
restriction formula.  More immediately, having explicit expressions for correlators 
in these theories will help to determine the $(0,2)$ mirror map needed to reconcile the results of Guffin and Katz~\cite{Guffin:2007mp}\ and those of~\cite{Adams:2003zy}. Finally, we expect very similar formulae to exist for $(0,2)$ theories that do not possess a $(2,2)$ locus. These include phenomenologically interesting cases; for example, the Landau-Ginzburg phase of the $(0,2)$ quintic with a rank $4$ bundle. 

\acknowledgments
We would like to thank Anirban Basu, Josh Guffin, Sheldon Katz and Ronen Plesser for useful discussions. This article is based upon work 
supported in part by the National Science Foundation under Grants 
PHY-0094328 and PHY-0401814 .  

\providecommand{\href}[2]{#2}\begingroup\raggedright\endgroup

\end{document}